\documentclass[twocolumn,showpacs,preprintnumbers,amsmath,amssymb]{revtex4}

\usepackage{graphicx}
\usepackage{dcolumn}
\usepackage{bm}

\begin{document}

\preprint{APS}

\title{Scaling functions for systems with finite range of interaction}

\author{C.I.N. Sampaio-Filho}
\email{cesar@fisica.ufc.br}
\affiliation{Departamento de F\'{i}sica, Universidade Federal do Cear\'{a} , 60451-970, Fortaleza-CE, Brasil}


\author{F.G.B. Moreira}
\email{brady@df.ufpe.br}
\affiliation{Departamento de F\'{i}sica Teorica e  Experimental, Universidade Federal do Rio Grande do Norte, 59072-970, Natal-RN, Brasil}



\date{\today}

\begin{abstract}
We present a numerical determination of the scaling functions of the magnetization, the susceptibility, and the  Binder's cumulant, for two nonequilibrium model systems with varying range of interactions. We consider Monte Carlo simulations of the block voter model (BVM) on square lattices and of the majority-vote model (MVM) on random graphs. In both cases, the satisfactory data collapse obtained for several system sizes and interaction ranges, supports the hypothesis that these functions are universal. Our analysis yields an accurate estimation of the long-range exponents, which govern the decay of the critical amplitudes with the range of interaction, and is consistent with the assumption that the static exponents are Ising-like for the BVM and classical for the MVM. 
\end{abstract}


                              
\maketitle


\section{\label{sec:level1}Introduction}
In statistical physics of equilibrium and nonequilibrium, the critical behavior characteristic of continuous order-disorder phase transitions is strongly dependent on the range of interactions.  Within the context of nonequilibrium phase transitions,  the influence of the range of interactions has been studied considering different models,  such as the contact process \cite{lubeck2003,ginelli2005,oliveira2007,odor2009}, models that present self-organized criticality \cite{gleiser2000,lubeck2004} and  the majority-vote  model (MVM) defined on regular  \cite{sampaio2011} and random networks \cite{brady2005,lima2008,brady2010}. The MVM  exhibits a continuous phase transition in a two-dimensional parameter space defined by the noise parameter $q$ (the probability that a spin adopts a state contrary of the state of the majority of its neighbors) and the strength of the range of the interaction $\Lambda$. A general conclusion from these studies   \cite{sampaio2011,brady2005,lima2008,brady2010} is that the transition occurs at a critical noise $q_c$ which is an increasing function of the parameter $\Lambda$. Moreover, it should also be emphasized that the critical amplitudes of relevant thermodynamical quantities become reduced as  the range of the interactions increases.

The range of interaction parameter, $\Lambda$, has a meaning which depends on the model system been studied. For instance, for inflow dynamics of spin systems \cite{sznajd2006} defined on regular lattices we may define $\Lambda=R_{eff}$ \cite{binder1993}, the maximum effective distance for the central spin be influenced by its neighbors. 
In a recent paper \cite{sampaio2011} we consider the collective behavior of the block voter model (BVM) which introduces long-ranged interactions in the system. The BVM is defined by an outflow dynamics where a central set of $N_{PCS}$ spins, denoted by persuasive cluster spins (PCS), tries to influence the opinion of their neighboring counterparts. It is shown that the effects of increasing the size of the persuasive cluster are the reduction of the critical amplitudes and the increment of the ordered region in the phase diagram \cite{sampaio2011}. Therefore, within the context of the present study, the range of interaction parameter $\Lambda$ is defined by the number of spins $N_{PCS}$ inside the persuasive cluster (that is, $\Lambda=N_{PCS}$). 
On the other hand, simulations of the MVM on classical random graphs  \cite{brady2005,brady2010} with varying mean connectivity $\kappa$ at fixed number of vertices $N$, show that the parameter $\kappa$ has a similar influence on both the phase diagram and the critical amplitudes of the relevant quantities.  Hence, in this case we have $\Lambda=\kappa$.

In the present study we perform Monte Carlo simulations of two nonequilibrium model systems, namely, the block voter model 
on the regular square lattice and the majority-vote model on random graph. Our main goal is to discuss and obtain the collapse 
of the magnetization, the susceptibility, and the Binder's fourth-order  cumulant, as well as their corresponding universal functions, including data from simulations of systems with different sizes $N$ and various values of the range of interaction parameter $\Lambda$. In Sec. II we introduce the finite-size scaling ansatz which also includes the range of interaction as a relevant scaling field.  Sec. III contains the results of the simulations and presents a discussion on how to determine the universal functions from the calculation of the usual static critical exponents and the new exponents related to the role played by the range of interaction parameter. We conclude in Sec. IV.

\section{\label{sec:level2} Finite-size scaling}

The finite-size scaling theory $(FSS)$ \cite{fisher1972,brezin1981,park2007,satorras2008} has been of great benefit to the understanding of numerical results of Monte Carlo simulations on finite systems. In this way, we perform the extrapolation to the thermodynamic limit ($N \rightarrow \infty$) in order to obtain reliable estimates of critical exponents and critical parameters. Moreover, the $FSS$ allows us to obtain universal functions representing the collapse of data for several values of $N$. 
For instance, the standard finite-size scaling equations for the order parameter, the susceptibility, and the Binder's fourth-order  cumulant are written as 

\begin{equation}
 M_{N}(q) \sim N^{-\beta/\overline{\nu}}\widetilde{M}(\varepsilon N^{1/\overline{\nu}}),
 \label{eq01}
\end{equation}
\begin{equation}
 \chi_{N}(q) \sim N^{\gamma/\overline{\nu}}\widetilde{\chi}(\varepsilon N^{1/\overline{\nu}}), 
 \label{eq02}
\end{equation}
\begin{equation}
 U_{N}(q) \sim \widetilde{U}(\varepsilon N^{1/\overline{\nu}}),
 \label{eq03}
\end{equation}
where $\epsilon = q - q_{c}$ is the distance from the critical noise parameter $q_{c}$. Note that for $d$-dimensional lattices with $N$ spins, $N=L^ d$ and $\overline{\nu}=d\nu$, where $\nu$ is the correlation length exponent. The exponents $\beta/ \overline{\nu}$ and $\gamma/\overline{\nu}$  are associated with the decay of the order parameter $M_{N}(q)$ and the divergence of the susceptibility $\chi_{N}(q)$,  respectively. The $\widetilde{M}$,  $\widetilde{\chi}$, and $\widetilde{U}$ are universal scaling functions of the scaling variable $\epsilon N^ {1/\overline{\nu}}$. 

We have mentioned that the presence of long-ranged interactions described by the parameter $\Lambda$ has influence on the nature of both  the phase diagram and  the critical fluctuations.  Yet  the above finite-size scaling equations  do not explain the decay of the critical amplitudes with the range of interaction \cite{sampaio2011,brady2005,lima2008,brady2010}.
In order to take into account this feature, we should add scale free terms in the parameter $\Lambda$, such that the new scaling relations for the relevant quantities are still generalized homogeneous functions. Therefore, in this paper we will consider the following ansatz for the scaling equations: 

\begin{equation}
M_{N}(q, \Lambda) = \Lambda^ {-X} N^{-\beta/\overline{\nu}} \widetilde{M}(\varepsilon N^{1/\overline{\nu}} \Lambda^{-Z}),
\label{eq04}
\end{equation}
\begin{equation}
 \chi_{N}(q, \Lambda) = \Lambda^ {-Y} N^{\gamma/\overline{\nu}}\widetilde{\chi}(\varepsilon N^{1/\overline{\nu}} \Lambda^{-Z}),
\label{eq05} 
\end{equation} 
\begin{equation}
u_{N}(q, \Lambda) = \Lambda^{-Z} N^{1/\overline{\nu}}\widetilde{u}(\varepsilon N^{1/\overline{\nu}} \Lambda^{-Z}),
\label{eq06}
\end{equation}
where $X$, $Y$, and $Z$ are, respectively,  nonnegative exponents associated with the critical amplitudes of the magnetization, of the susceptibility \cite{binder1993}, and of the derivative of the Binder's cumulant $u_{N}(q, \Lambda) = \frac{dU}{dq}$. The minus signs in the respective power laws are consistent with the decay of the critical amplitudes with the parameter $\Lambda$.  

By definition, the critical amplitude of the cumulant $U_{N}$ does not depend on the size of the system \cite{binder1981}. Considering the derivative of (\ref{eq03}) with respect to the noise parameter $q$, the resulting equation give us an expression where the critical amplitude is size dependent.
However,  it is possible to show that the  critical amplitude of the derivate $u_{N}(q, \Lambda)$ is also dependent on the range of interaction (see Eq. (\ref{eq06})).   In order to have this feature into account we have introduced a new scaling variable $\eta=\varepsilon N^{1/\overline{\nu}} \Lambda^{-Z}$, which also incorporates the range of interaction parameter $\Lambda$ in its definition.

In the next section we will demonstrate the scaling equations (\ref{eq04}), (\ref{eq05}), and (\ref{eq06}), by performing Monte Carlo simulations for two distinct model systems defined on square lattices and random graphs. In particular, we will show how to obtain the universal functions $\widetilde{M}$,  $\widetilde{\chi}$, and $\widetilde{U}$.

\section{Monte Carlo simulations}
\subsection{Regular Lattice}

The block voter model is a nonequilibrium model defined by an outflow dynamics \cite{sznajd2006} where a central set of $N_{PCS}$ spins, denoted by persuasive cluster spins, tries to influence the opinion of their neighboring counterparts. For $N_{PCS}>2$, the system exhibits an order-disorder phase transition at a critical noise parameter $q_{c}$, which is a monotonically increasing function of the size of the persuasive cluster.  For finite size of $N_{PCS}$ the critical behavior is given by the Ising universality class. Shortly, the $BVM$ has the same properties of the majority-vote model \cite{oliveira1992}, but considering outflow dynamics and introducing the parameter $N_{PCS}$ which increases the region of the ordered phase in the phase diagram \cite{sampaio2011}.

For our purpose it is necessary an accurate determination of the critical noise parameter $q_{c}$ for the values of $N_{PCS}=9, 16, 25, 36, 49, 64$ considered. The phase diagram of the model in the  $q-N_{PCS}$ parameter space was reported in \cite{sampaio2011}. From the results for $q_{c}$, we simulate the $BVM$ on regular square lattices of linear length $L = 100, 160, 180, 200$, and $300$, considering periodic boundary conditions and asynchronous update. Therefore, $1$ Monte Carlo step ($MCS$) is accomplished by repeating the following procedure $N$ times: choose randomly one adjacent site of the persuasive cluster and try to flip it with the probability tax given by
\begin{equation}
 w(\sigma_{i})=\frac{1}{2}\left[1-(1-2q)\sigma_{i}S(\sum_{\delta = 1}^{\Lambda}  \sigma_{i+\delta}) \right],
 \label{eq13}
\end{equation}
where the summation is over all $\Lambda=N_{PCS}$ sites that make up the persuasive cluster, and $S(x) = sgn(x)$ if $x\neq 0$ and otherwise $S(0)=0$. As all analysis is made at the critical region, we wait $3\times 10^{4}$ $MCS$ to make the system to reach the steady state and the time averages are estimated during the next $40\times 10^{4}$ $MCS$.  
For	all	sets	of	parameters	$(q ,N_{PCS} )$,	at least $100$ independent runs (samples) were considered in the calculation of the configurational averages. The simulations were performed using different initial spin configurations. We have checked that the numerical results do not depend on the initial fraction of spins in the state $\sigma = 1$.

\begin{figure}[htb]
\begin{center}
\includegraphics[width=8.0cm,height=6.5cm]{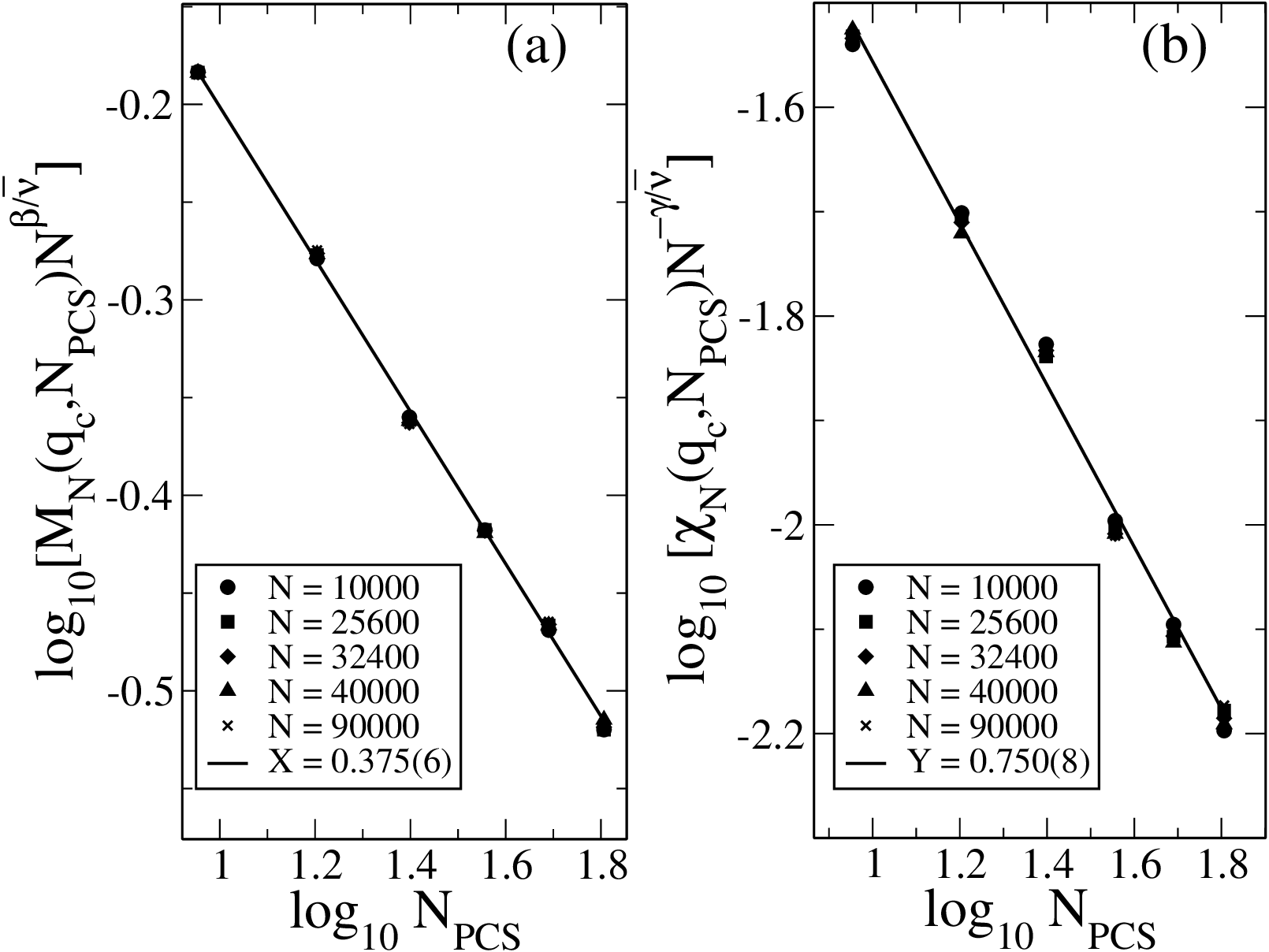}
\caption{\label{fig1}The estimation for the exponents X and Y.  The dependence on $N_{PCS}$  of (a) the magnetization   and  (b) the susceptibility, measured at $q_{c}$.  Each point is averaged over five lattice sizes.The straight lines  represent the scaling relations $M_{N} \sim N_{PCS}^{-X}$ and $\chi_{L} \sim N_{PCS}^{-Y}$, whose slopes yield $X = 0.375(6)$  and $Y = 0.750(8)$.}
\end{center}
\end{figure}

Firstly we calculate the exponents $X$ and $Y$ associated with the critical amplitudes of the magnetization and susceptibility.  Considering $\Lambda=N_{PCS}$ and multiplying  the Eqs. (\ref{eq04}) and (\ref{eq05}),  respectively,  by $N^{\beta/\overline{\nu}}$ and $N^{-\gamma/\overline{\nu}}$,  we obtain the results shown in Fig. \ref{fig1}. 
 In this  $log$-$log$ plot we show the critical magnetization (Fig. 1a) and the critical susceptibility (Fig. 1b) {\it{versus}} $N_{PCS}$.   We consider $\beta/\nu=0.125$, $\gamma/\nu=1.75$, and $\nu = 1$ ($\overline{\nu}=2$), which are the non-classical exponents for the BVM  on the square lattice \cite{sampaio2011}. For every $N_{PCS}$, we have five values of $M_{L}$ and $\chi_{L}$ that are associated with the sizes of the lattices considered. A linear regression of this set of points yields $X=0.375(6)$ and $Y=0.750(8)$. 

The exact values of these exponents can be determined from the following relations:
\begin{equation}
X = \frac{\beta_{MF} - \beta}{2\phi},
\label{eq08}
\end{equation}
\begin{equation}
Y = \frac{\gamma - \gamma_{MF}}{2\phi},
\label{eq09}
\end{equation}
where   $\phi = \nu_{MF}(d_{c} - d)/d$ is the crossover exponent \cite{ginzburg1960, binder1993, binder1996, lubeck2003}. 
The above equations were obtained within the context of the crossover from  non-mean-field to classical scaling behavior   \cite{lubeck2003,lubeck2003B}.
Taking into account  the Ising exponents, $\beta=0.125$ and $\gamma=1.75$, the corresponding values of the classical exponents, e.g., $\beta_{MF} = 0.5$, $\gamma_{MF}=1.0$, and $\nu_{MF}=0.5$, and the upper critical dimension $d_{c}=4$ of the block voter model, we have $X=0.375$ and $Y=0.750$. Therefore, the numerical results are in good accordance with the exact values.

\begin{figure}[htb]
\begin{center}
\includegraphics[width=8.0cm,height=6.5cm]{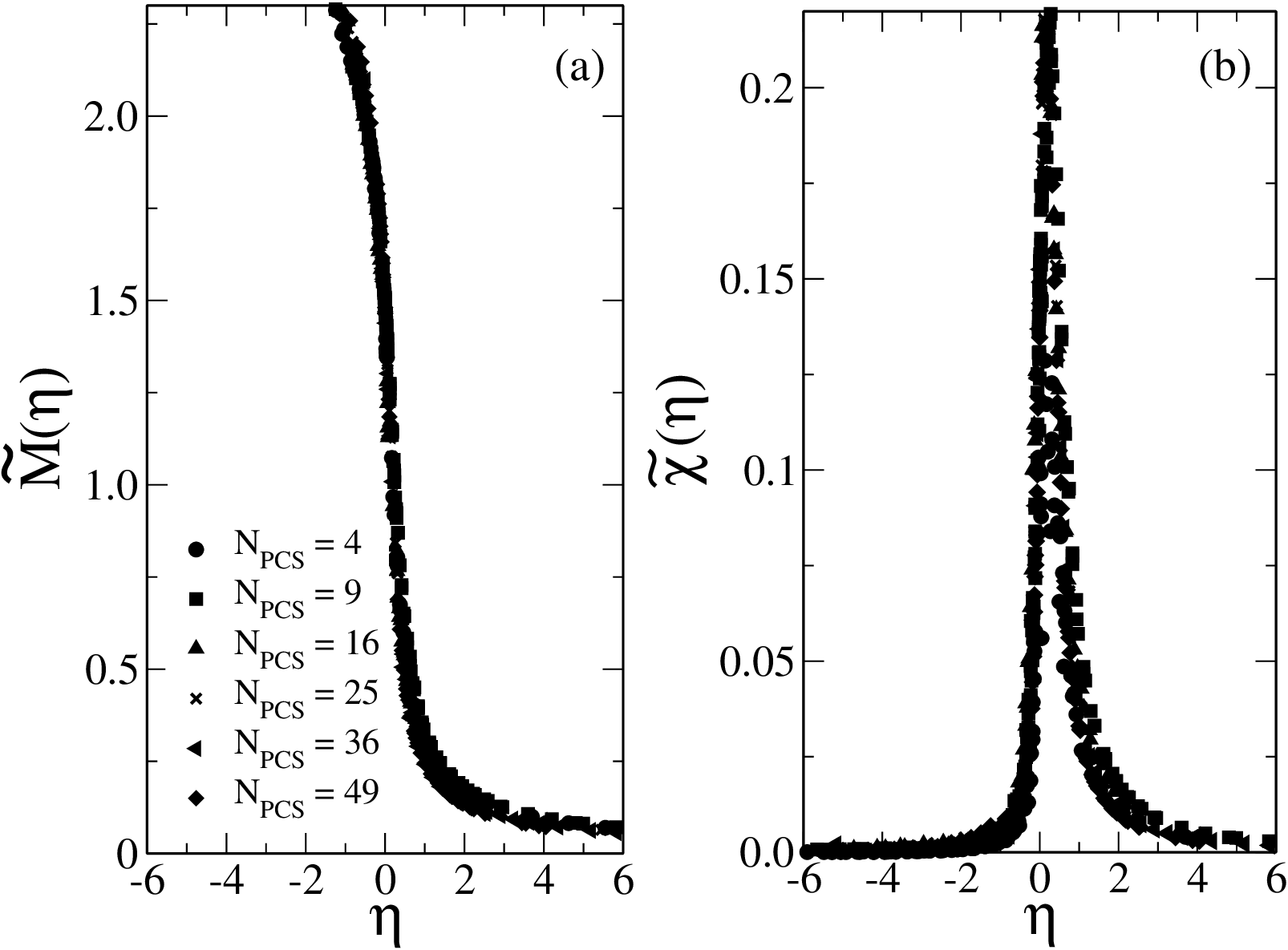}
\caption{\label{fig02} Data collapse of the order parameter (a) and of the susceptibility (b) for $N_{PCS}=4, 9,16, 25, 36, 49$. For each value of $N_{PCS}$ we have systems  of sizes $N = 10000, 25600, 32400, 40000, 90000$. The universal functions are consistent with Ising exponents: $\beta/\nu = 0.125$,  $\gamma/\nu = 1.75$, and $\nu = 1.0$. We use $X=0.375$, $Y=0.750$, and $Z=0.250$ for the long-range exponents of the block voter model.}
\end{center}
\end{figure}

\begin{figure}[b]
\begin{center}
\includegraphics[width=8.5cm,height=6.5cm]{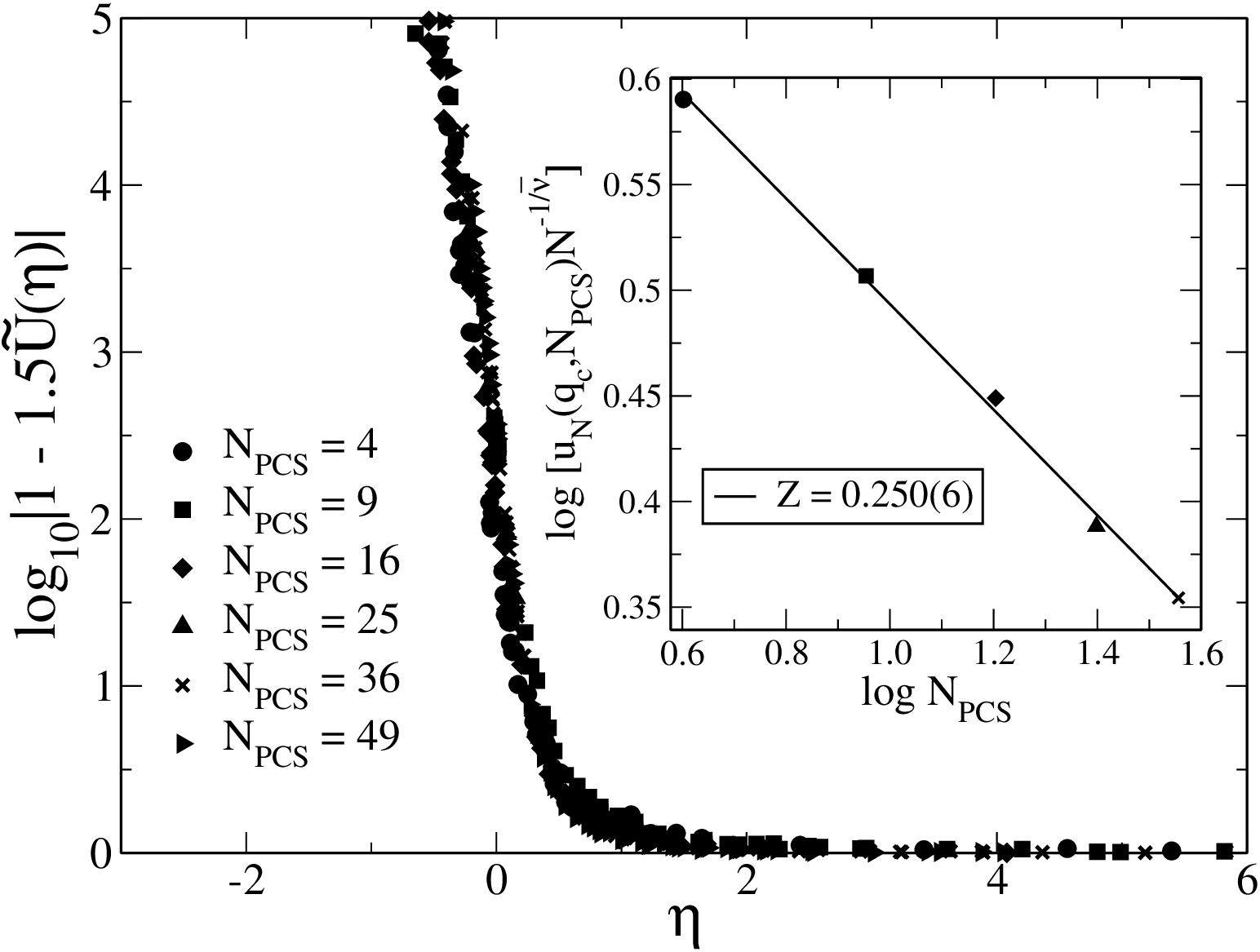}
\caption{\label{fig03} The universal function $\widetilde{U}(\eta)$, where $\eta = \varepsilon N^{1/\overline{\nu}} N_{PCS}^{-Z}$. The data collapse includes data for six different values of  $N_{PCS}$ and  $N = 10000, 25600, 32400, 40000, 90000$. The set of exponents is the same as in Fig. 2. The inset illustrates the method used for obtaining the exponent $Z$.}
\end{center}
\end{figure}

We now apply the method described above to evaluate the exponent $Z$. Multiplying  the Eq. (\ref{eq06}) by $N^{-1/\overline{\nu}}$, the critical amplitude of  $u_{N}(q,N_{PCS} )N^{-1/\overline{\nu}}$ varies as a power law of the size of the persuasive cluster spin, with exponent $Z$. In the inset of Fig.  (\ref{fig03}) we plot this quantity  as a function of  $N_{PCS}$, where $\overline{\nu}=2$ and each point represents the average over five different system sizes. A linear regression of this set of points yields $Z = 0.250(6)$.

The existence of the universal scaling functions $\widetilde{M}(\eta) = M_{N}(q, N_{PCS})N^{\beta/\overline{\nu}}N_{PCS}^{X}$, $\widetilde{\chi}(\eta) = \chi_{N}(q, N_{PCS})N^{-\gamma/\overline{\nu}}N_{PCS}^{Y}$, and $\widetilde{U}(\eta) = U_{N}(q, N_{PCS})$, where the scaling variable is defined as $\eta = \varepsilon N^{1/\overline{\nu}}N_{PCS}^{-Z}$, suggests that the data point of the corresponding quantity obtained from simulations with different values of $N$ and $N_{PCS}$ should collapse into a single universal curve.  Figs. 2 and 3 show the data collapse for the order parameter, the susceptibility and the Binder's cumulant, considering five values of the system size $N$ and six values of the number of persuasive spins $N_{PCS}$. We use the following set of exponents: $\beta/\nu = 0.125$, $\gamma/\nu = 1.75$, $1/\nu = 1.0$, $X = 0.375$, $Y = 0.750$, and $Z = 0.250$.  It is worth mentioning that only with the correct exponents the universal curves are obtained. The resulting satisfactory collapses  show not only that the previously calculated values for the exponents associated with the interaction range are correct, but also verify the validity of the ansatz defined  by Eqs. (\ref{eq04}), (\ref{eq05}), and (\ref{eq06}).

\subsection{\label{sec:level3}Random Network}

In this subsection we consider  the analysis of the continuous phase transitions of the majority-vote model (MVM) on classical random graphs \cite{albert2002,newman2003,latora2006,mendes2008}. As can be noticed in references \cite{brady2005} and \cite{brady2010}, the effect of varying the average degree of the random graph is to increase the ordered region in the phase diagram and to reduce the critical fluctuations. A similar feature was observed with respect to the role played by the long-range parameter $N_{PCS}$ in the previous analysis of the block voter model on square lattices. We therefore conjecture that  the finite-size scaling ansatz (Eqs. $4-6$) can be used to obtain the universal functions for the MVM on random graphs, once the long-range parameter $\Lambda$ is replaced by the average degree $\kappa$. 

We perform Monte Carlo simulations of the majority-vote model with noise in the case where each spin is associated with a vertex of an Erdos-Renyi random graph and can have the values $\pm 1$. The two-state majority-vote model is a nonequilibrium model defined by an inflow dynamics \cite{sznajd2006} where a central spin  agrees with the state of the majority of its neighbors, with probability $1-q$, and it disagrees with probability $q$. The phase diagram of the model in the entire  $q-\kappa$ parameter space was reported in \cite{brady2005}.
Here we consider the results from simulations of graphs of sizes  $N = 8000, 10000, 15000, 20000, 40000$, and, for each value of $N$, varied connectivity $k=4, 6, 8, 10, 20, 30$. We employ the configuration method \cite{mendes2008}  to generate classical random networks. We use asynchronous update and $1$ Monte Carlo step ($MCS$) corresponds to $N$ tries of flipping a randomly chosen spin according to the rule (\ref{eq13}). Typically we wait  $3\times 10^{4}$ $MCS$ to make the system reach the steady state and the time averages are estimated considering the next $40\times 10^{4}$ $MCS$. 
For	each set of parameters $(q ,\kappa )$,	we generate $100$ independent samples in order to calculate the configurational averages. We have checked that the numerical results do not depend on the initial spin configurations, that is, on the initial fraction of spins in the state $\sigma = 1.$

\begin{figure}[htb]
\begin{center}
\includegraphics[width=8.0cm,height=6.5cm]{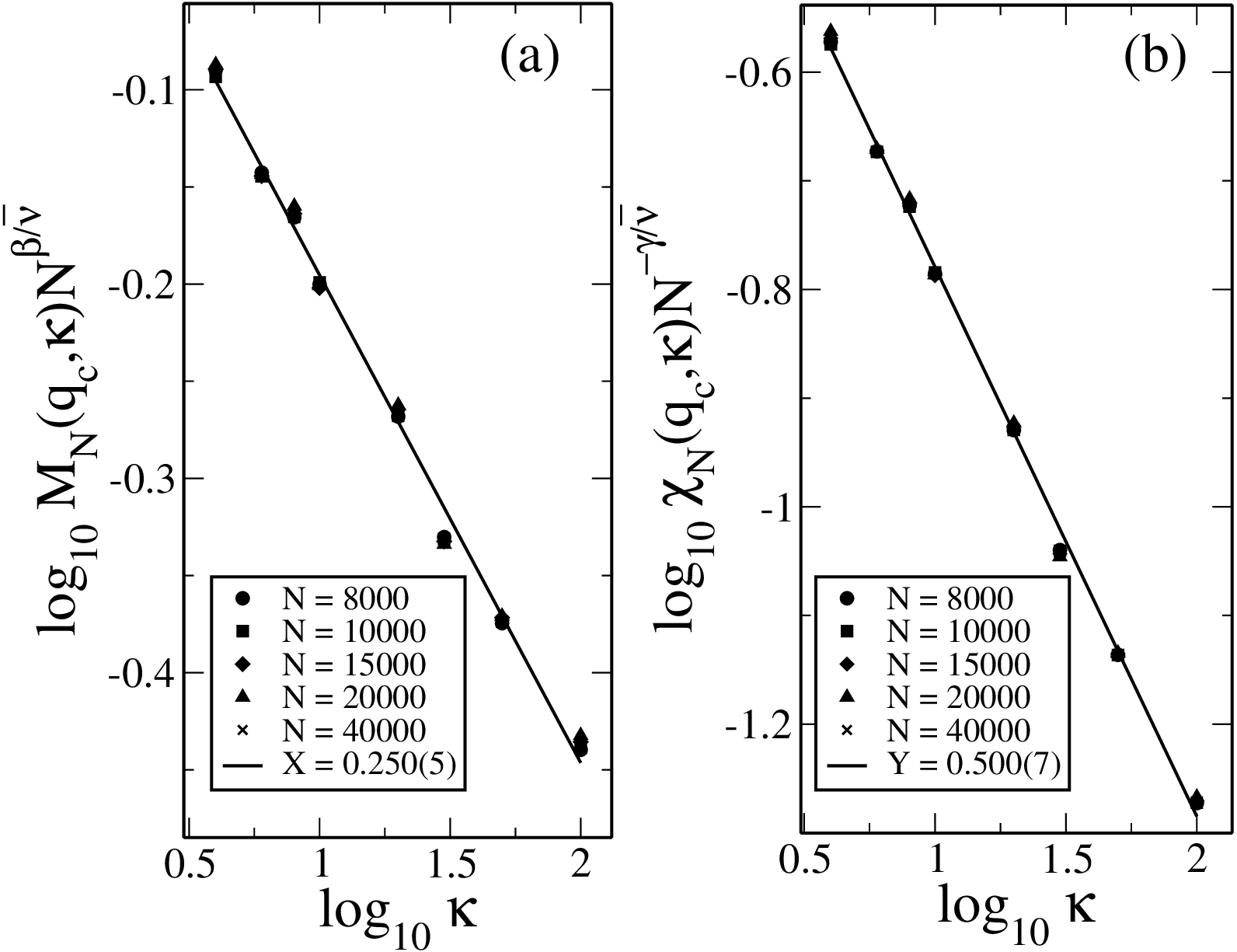}
\caption{\label{fig04}The estimation for the exponents X and Y. (a) Plot of $M_{N}$ measured at $q_{c}$ against $\kappa$. The solid line represents the relation $M_{N} \sim \kappa^{-X}$ with $X = 0.250(5)$. (b) Plot of the critical susceptibility $\chi_{N}$ against $\kappa$.  The solid line represents the relation $\chi_{N} \sim \kappa^{-Y}$ with $Y = 0.500(7)$. Each point corresponds to the average over five values of $N$.}
\end{center}
\end{figure}

The calculation of the long-range exponents $X, Y,$ and $Z$ of the majority-vote model on random graphs follows exactly the same procedure as that used for the block voter model on square lattices. Fig. \ref{fig04} shows the data for the critical magnetization and the critical susceptibility as functions of $\kappa$, obtained from simulations of systems with five different sizes. Considering  the mean-field classical exponents,  $\beta_{MF} = 0.5$, $\gamma_{MF}=1.0$,  $\nu_{MF}=0.5$, and the upper critical dimension $d_{c}=4$, a linear regression of the data points in Fig. 4(a) and Fig 4(b) yields $X=0.250(5)$ and $Y=0.500(7)$, respectively. 
Even though these numerical results are consistent with the relation $Y=2X$, it should be noted that the Eqs. (\ref{eq08}) and (\ref{eq09}) do not apply to the MVM on random graph. For the exponent $Z$, see the inset of Fig. \ref{fig06},  we obtained  $Z=0.125(2)$.

\begin{figure}[htb]
\begin{center}
\includegraphics[width=8.5cm,height=6.5cm]{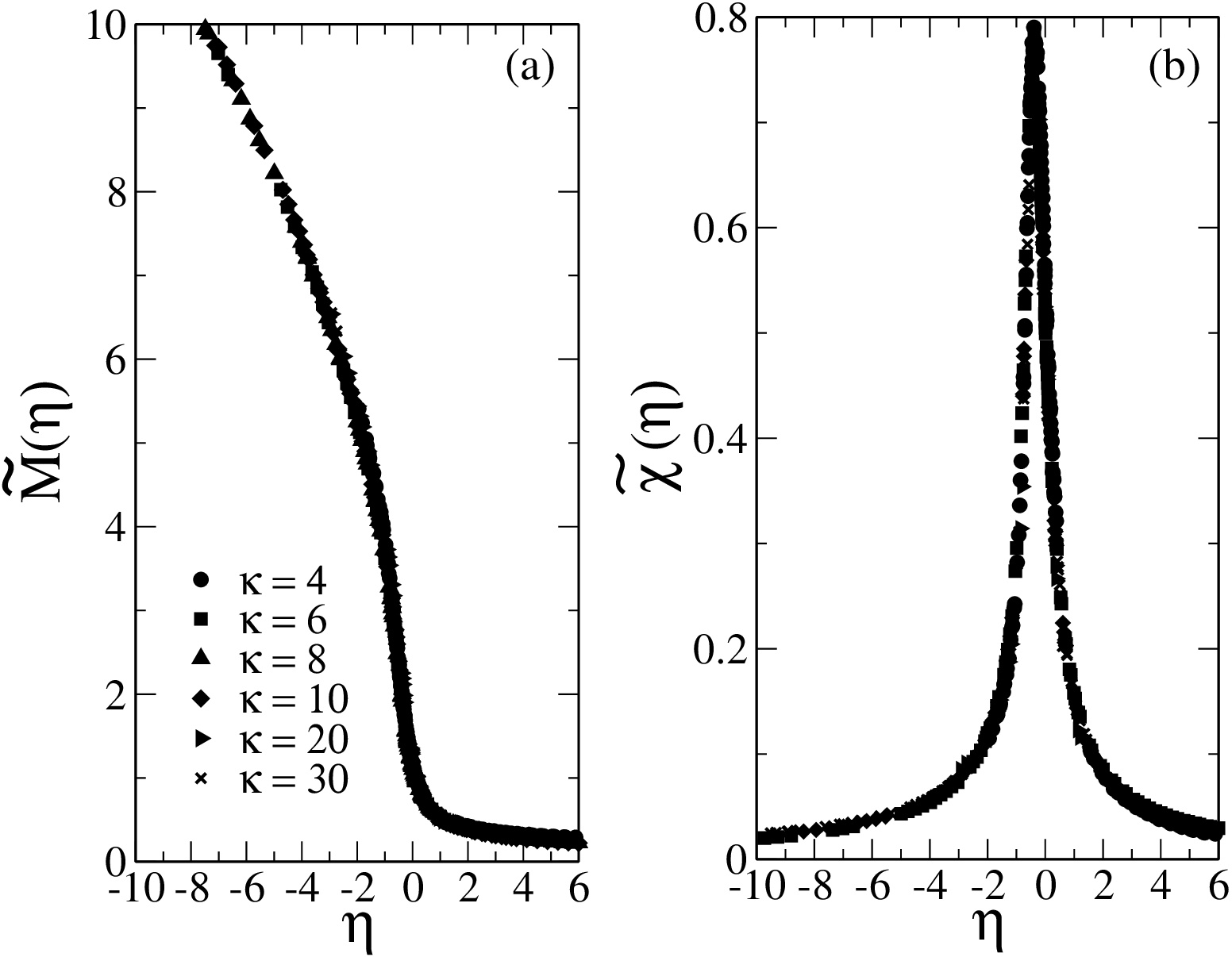}
\caption{\label{fig05} The universal functions $\widetilde{M}(\eta)$ and $\widetilde{\chi}(\eta)$, where $\eta = \varepsilon N^{1/\overline{\nu}}\kappa^{-Z}$, for the MVM on random graphs. To obtain the data collapse for $\kappa=4, 6, 8, 10, 30$ and  $N = 8000, 10000, 15000, 20000, 40000$  we use the exponents: $\beta = 0.5$,  $\gamma = 1.0$, $\nu = 0.5$, $X=0.25$, $Y=0.50$, $Z=0.125$.}
\end{center}
\end{figure}

\begin{figure}[b]
\begin{center} 
\includegraphics[width=8.5cm,height=6.5cm]{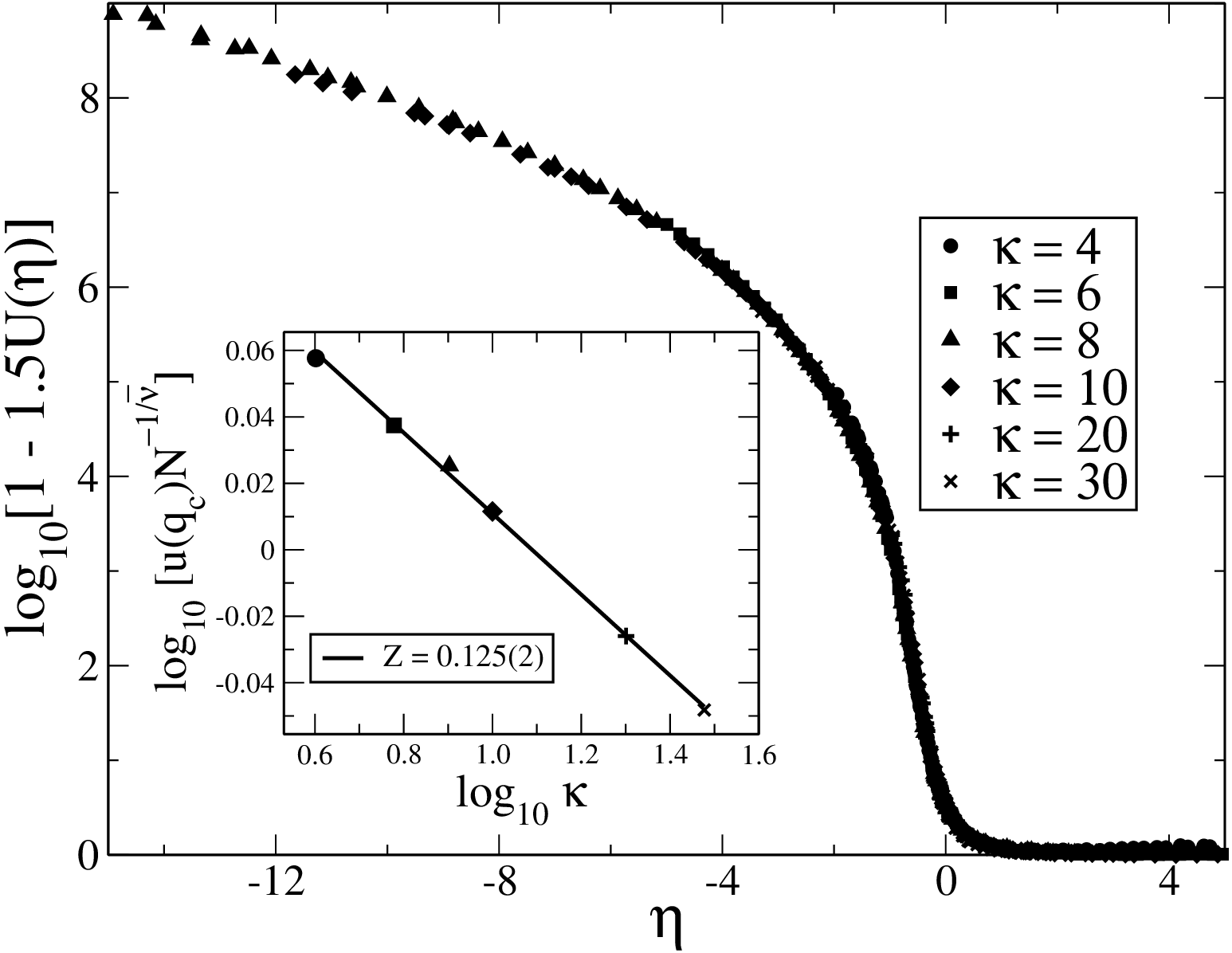}
\caption{\label{fig06} The data collapse of the Binder's cumulant, $\widetilde{U}(\eta)$,  for six values of the mean connectivity $\kappa$ and  $N = 8000, 10000, 15000, 20000, 40000$. The set of exponents is the same as in Fig. 5. The inset shows the evaluation of the exponent $Z$.}
\end{center}
\end{figure}

In Figs. \ref{fig05} and \ref{fig06} we show the universal functions $\widetilde{M}(\eta) = M_{N}(q, \kappa)N^{\beta/\overline{\nu}} \kappa^{X}$, $\widetilde{\chi}(\eta) = \chi_{N}(q, \kappa)N^{-\gamma/\overline{\nu}}\kappa^{Y}$, and $\widetilde{U}(\eta) = U_{N}(q, \kappa)$, where $\eta = \varepsilon N^{1/\overline{\nu}}\kappa^{-Z}$. Once again we emphasize that the good quality of these data collapses, which result from simulations of systems with  five different sizes $N$ and six values for the average connectivity $\kappa$, is a strong evidence in favor of the scaling ansatz  (\ref{eq04} - \ref{eq06}), as well as of using the correct set of exponents.

\section{\label{sec:level3}Conclusion}

In this work we investigated the effects of long-ranged interactions on the critical amplitudes of the magnetization, the susceptibility, and the Binder's cumulant, for two dynamical systems defined on regular square lattices and random networks. Our results from Monte Carlo simulations of systems with different sizes, $N$,  and varying range of interaction, $\Lambda$, were analyzed through a finite-size scaling ansatz which defines universal functions of a single scaling variable $\eta=\varepsilon N^{1/\overline{\nu}} \Lambda^{-Z}$ and introduces new exponents $X$, $Y$, and $Z$, governing the decay of the  critical amplitudes with the long-range parameter $\Lambda$, besides the static exponents $\beta$, $\gamma$, and  $\nu$, describing the dependence with $N$ of the calculated quantities. 

From the data collapse of the numerical results, we succeeded in determining the  universal scaling functions. For the block voter model on square lattices, a nonequilibrium system in the universality class of the equilibrium two-dimensional Ising model \cite{sampaio2011}, the resulting collapses were obtained using the exact values of the static exponents, $\beta = 0.125$, $\gamma = 1.75$, $\nu = 1.0$, whereas our estimation of the long-range exponents yields $X=0.375$ and $Y=0.750$,  in agreement  with available exact results \cite{lubeck2003},  and $Z=0.250$.  It is worth mentioning that this is the first determination of the exponent $Z$ for a system in the Ising universality class. 

For the majority-vote model on random graphs, the universal functions are consistent  with classical mean-field exponents, $\beta = 0.5$,  $\gamma = 1.0$, $\nu = 0.5$, whereas the quoted values  $X=0.250$, $Y=0.500$, and $Z=0.125$ represent the first calculation of the long-range exponents for a model defined on a random network.  The present conclusion in favor of classical exponents is in disagreement with the work of Pereira and Moreira \cite{brady2005}, which reported that the exponents of the MVM on random graphs are different from the classical mean-field exponents. For a given model system, we should expect to obtain the same critical exponents regardless of  whether the interaction range is considered in the scaling functions. In particular, 
the exponents $\beta$, $\gamma$, and $\nu$, determining the system-size dependence in the critical region, are not affected by details such as the interaction range, since their values only depend  on universality arguments. Therefore the observed difference between the present results and those of \cite{brady2005}  are not attributed to the inclusion of the interaction range in the scaling ansatz. In fact,  the linear behavior shown in Figs. \ref{fig04} and \ref{fig06} is only observed by using the correct set of classical exponents reported here.

In summary, the quite good data collapse obtained from simulations of two nonequilibrium model systems including results for several system sizes and a large range of the interaction, strongly supports the finite-size scaling ansatz defined by Eqs. $(4-6)$. It would be therefore  well worth to extend the present study to regular lattices in higher dimension, $d=3$ for example, as well as to other sort of complex networks. In the latter case an interesting question arises about the identification of the corresponding long-ranged parameter $\Lambda$.

\begin{acknowledgments}
C.I.N. Sampaio-Filho is supported by Coordena\c{c}\~{a}o de Aperfei\c{c}oamento de Pessoal de N\'{i}vel Superior (CAPES) and Funda\c{c}\~{a}o Cearense de Apoio ao Desenvolvimento Cient\'{i}fico e Tecnol\'{o}gico (Funcap). FGBM thanks  Funda\c{c}\~{a}o Norte-Rio-Grandense de Pesquisa e Cultura (FUNPEC) and Instituto Internacional de F\'{i}sica/UFRN for a research grant.
\end{acknowledgments}


\bibliography{scaling}

\providecommand{\noopsort}[1]{}\providecommand{\singleletter}[1]{#1}%
\begin{thebibliography}{25}
\expandafter\ifx\csname natexlab\endcsname\relax\def\natexlab#1{#1}\fi
\expandafter\ifx\csname bibnamefont\endcsname\relax
  \def\bibnamefont#1{#1}\fi
\expandafter\ifx\csname bibfnamefont\endcsname\relax
  \def\bibfnamefont#1{#1}\fi
\expandafter\ifx\csname citenamefont\endcsname\relax
  \def\citenamefont#1{#1}\fi
\expandafter\ifx\csname url\endcsname\relax
  \def\url#1{\texttt{#1}}\fi
\expandafter\ifx\csname urlprefix\endcsname\relax\def\urlprefix{URL }\fi
\providecommand{\bibinfo}[2]{#2}
\providecommand{\eprint}[2][]{\url{#2}}

\bibitem[{\citenamefont{Lubeck}(2003)}]{lubeck2003}
\bibinfo{author}{\bibfnamefont{S.}~\bibnamefont{Lubeck}},
  \bibinfo{journal}{Phys. Rev. Lett.} \textbf{\bibinfo{volume}{90}},
  \bibinfo{pages}{210601} (\bibinfo{year}{2003}).

\bibitem[{\citenamefont{Ginelli et~al.}(2005)\citenamefont{Ginelli, Hinrichsen,
  Livi, Mukamel, and Politi}}]{ginelli2005}
\bibinfo{author}{\bibfnamefont{F.}~\bibnamefont{Ginelli}},
  \bibinfo{author}{\bibfnamefont{H.}~\bibnamefont{Hinrichsen}},
  \bibinfo{author}{\bibfnamefont{R.}~\bibnamefont{Livi}},
  \bibinfo{author}{\bibfnamefont{D.}~\bibnamefont{Mukamel}}, \bibnamefont{and}
  \bibinfo{author}{\bibfnamefont{A.}~\bibnamefont{Politi}},
  \bibinfo{journal}{Phys. Rev. E} \textbf{\bibinfo{volume}{71}},
  \bibinfo{pages}{026121} (\bibinfo{year}{2005}).

\bibitem[{\citenamefont{Fiore and de~Oliveira}(2007)}]{oliveira2007}
\bibinfo{author}{\bibfnamefont{C.~E.} \bibnamefont{Fiore}} \bibnamefont{and}
  \bibinfo{author}{\bibfnamefont{M.~J.} \bibnamefont{de~Oliveira}},
  \bibinfo{journal}{Phys. Rev. E.} \textbf{\bibinfo{volume}{76}},
  \bibinfo{pages}{041103} (\bibinfo{year}{2007}).

\bibitem[{\citenamefont{Juh\'{a}sz and \'{O}dor}(2009)}]{odor2009}
\bibinfo{author}{\bibfnamefont{R.}~\bibnamefont{Juh\'{a}sz}} \bibnamefont{and}
  \bibinfo{author}{\bibfnamefont{G.}~\bibnamefont{\'{O}dor}},
  \bibinfo{journal}{Phys. Rev. E} \textbf{\bibinfo{volume}{80}},
  \bibinfo{pages}{041123} (\bibinfo{year}{2009}).

\bibitem[{\citenamefont{Gleiser et~al.}(2000)\citenamefont{Gleiser, Tamarit,
  and Cannas}}]{gleiser2000}
\bibinfo{author}{\bibfnamefont{P.~M.} \bibnamefont{Gleiser}},
  \bibinfo{author}{\bibfnamefont{F.~A.} \bibnamefont{Tamarit}},
  \bibnamefont{and} \bibinfo{author}{\bibfnamefont{S.~A.}
  \bibnamefont{Cannas}}, \bibinfo{journal}{Physica A: Statistical Mechanics and
  its Applications} \textbf{\bibinfo{volume}{275}}, \bibinfo{pages}{272 }
  (\bibinfo{year}{2000}), ISSN \bibinfo{issn}{0378-4371}.

\bibitem[{\citenamefont{Lubeck}(2004)}]{lubeck2004}
\bibinfo{author}{\bibfnamefont{S.}~\bibnamefont{Lubeck}},
  \bibinfo{journal}{Phys. Rev. E.} \textbf{\bibinfo{volume}{69}},
  \bibinfo{pages}{066101} (\bibinfo{year}{2004}).

\bibitem[{\citenamefont{Sampaio-Filho and Moreira}(2011)}]{sampaio2011}
\bibinfo{author}{\bibfnamefont{C.~I.~N.} \bibnamefont{Sampaio-Filho}}
  \bibnamefont{and} \bibinfo{author}{\bibfnamefont{F.~G.~B.}
  \bibnamefont{Moreira}}, \bibinfo{journal}{Phys. Rev. E}
  \textbf{\bibinfo{volume}{84}}, \bibinfo{pages}{051133}
  (\bibinfo{year}{2011}).

\bibitem[{\citenamefont{Pereira and Moreira}(2005)}]{brady2005}
\bibinfo{author}{\bibfnamefont{L.~F.~C.} \bibnamefont{Pereira}}
  \bibnamefont{and} \bibinfo{author}{\bibfnamefont{F.~G.~B.}
  \bibnamefont{Moreira}}, \bibinfo{journal}{Phys. Rev. E}
  \textbf{\bibinfo{volume}{71}}, \bibinfo{pages}{016123}
  (\bibinfo{year}{2005}).

\bibitem[{\citenamefont{Lima et~al.}(2008)\citenamefont{Lima, Sousa, and
  Summor}}]{lima2008}
\bibinfo{author}{\bibfnamefont{F.~W.~S.} \bibnamefont{Lima}},
  \bibinfo{author}{\bibfnamefont{A.~O.} \bibnamefont{Sousa}}, \bibnamefont{and}
  \bibinfo{author}{\bibfnamefont{M.~A.} \bibnamefont{Summor}},
  \bibinfo{journal}{Physica A} \textbf{\bibinfo{volume}{387}},
  \bibinfo{pages}{3503} (\bibinfo{year}{2008}).

\bibitem[{\citenamefont{Melo et~al.}(2010)\citenamefont{Melo, Pereira, and
  Moreira}}]{brady2010}
\bibinfo{author}{\bibfnamefont{D.~F.~F.} \bibnamefont{Melo}},
  \bibinfo{author}{\bibfnamefont{L.~F.~C.} \bibnamefont{Pereira}},
  \bibnamefont{and} \bibinfo{author}{\bibfnamefont{F.~G.~B.}
  \bibnamefont{Moreira}}, \bibinfo{journal}{J. Stat. Phys.}
  \textbf{\bibinfo{volume}{10}}, \bibinfo{pages}{11032} (\bibinfo{year}{2010}).

\bibitem[{\citenamefont{Sznajd-Weron and Krupa}(2006)}]{sznajd2006}
\bibinfo{author}{\bibfnamefont{K.}~\bibnamefont{Sznajd-Weron}}
  \bibnamefont{and} \bibinfo{author}{\bibfnamefont{S.}~\bibnamefont{Krupa}},
  \bibinfo{journal}{Phys. Rev E.} \textbf{\bibinfo{volume}{74}},
  \bibinfo{pages}{031109} (\bibinfo{year}{2006}).

\bibitem[{\citenamefont{Mon and Binder}(1993)}]{binder1993}
\bibinfo{author}{\bibfnamefont{K.~K.} \bibnamefont{Mon}} \bibnamefont{and}
  \bibinfo{author}{\bibfnamefont{K.}~\bibnamefont{Binder}},
  \bibinfo{journal}{Phys. Rev. E} \textbf{\bibinfo{volume}{48}},
  \bibinfo{pages}{2498} (\bibinfo{year}{1993}).

\bibitem[{\citenamefont{Fisher and Barber}(1972)}]{fisher1972}
\bibinfo{author}{\bibfnamefont{M.~E.} \bibnamefont{Fisher}} \bibnamefont{and}
  \bibinfo{author}{\bibfnamefont{M.~N.} \bibnamefont{Barber}},
  \bibinfo{journal}{Phys. Rev. Lett.} \textbf{\bibinfo{volume}{28}},
  \bibinfo{pages}{1516} (\bibinfo{year}{1972}).

\bibitem[{\citenamefont{Brezin}(1982)}]{brezin1981}
\bibinfo{author}{\bibfnamefont{E.}~\bibnamefont{Brezin}}, \bibinfo{journal}{J.
  Physique} \textbf{\bibinfo{volume}{43}}, \bibinfo{pages}{15}
  (\bibinfo{year}{1982}).

\bibitem[{\citenamefont{Hong et~al.}(2007)\citenamefont{Hong, Ha, and
  Park}}]{park2007}
\bibinfo{author}{\bibfnamefont{H.}~\bibnamefont{Hong}},
  \bibinfo{author}{\bibfnamefont{M.}~\bibnamefont{Ha}}, \bibnamefont{and}
  \bibinfo{author}{\bibfnamefont{H.}~\bibnamefont{Park}},
  \bibinfo{journal}{Phys. Rev. Lett.} \textbf{\bibinfo{volume}{98}},
  \bibinfo{pages}{258701} (\bibinfo{year}{2007}).

\bibitem[{\citenamefont{Castellano and Pastor-Satorras}(2008)}]{satorras2008}
\bibinfo{author}{\bibfnamefont{C.}~\bibnamefont{Castellano}} \bibnamefont{and}
  \bibinfo{author}{\bibfnamefont{R.}~\bibnamefont{Pastor-Satorras}},
  \bibinfo{journal}{Phys. Rev. Lett.} \textbf{\bibinfo{volume}{100}},
  \bibinfo{pages}{148701} (\bibinfo{year}{2008}).

\bibitem[{\citenamefont{Binder}(1981)}]{binder1981}
\bibinfo{author}{\bibfnamefont{K.}~\bibnamefont{Binder}}, \bibinfo{journal}{Z.
  Phys. B} \textbf{\bibinfo{volume}{43}}, \bibinfo{pages}{119}
  (\bibinfo{year}{1981}).

\bibitem[{\citenamefont{de~Oliveira}(1992)}]{oliveira1992}
\bibinfo{author}{\bibfnamefont{M.~J.} \bibnamefont{de~Oliveira}},
  \bibinfo{journal}{J. Stat. Phys.} \textbf{\bibinfo{volume}{66}},
  \bibinfo{pages}{273} (\bibinfo{year}{1992}).

\bibitem[{\citenamefont{Ginzburg}(1960)}]{ginzburg1960}
\bibinfo{author}{\bibfnamefont{V.~L.} \bibnamefont{Ginzburg}},
  \bibinfo{journal}{Sov. Phys. Solid State} \textbf{\bibinfo{volume}{2}},
  \bibinfo{pages}{1824} (\bibinfo{year}{1960}).

\bibitem[{\citenamefont{Luijten et~al.}(1996)\citenamefont{Luijten, Blote, and
  Binder}}]{binder1996}
\bibinfo{author}{\bibfnamefont{E.}~\bibnamefont{Luijten}},
  \bibinfo{author}{\bibfnamefont{H.~W.~J.} \bibnamefont{Blote}},
  \bibnamefont{and} \bibinfo{author}{\bibfnamefont{K.}~\bibnamefont{Binder}},
  \bibinfo{journal}{Phys. Rev. E} \textbf{\bibinfo{volume}{54}},
  \bibinfo{pages}{4626} (\bibinfo{year}{1996}).

\bibitem[{\citenamefont{Lubeck and Heger}(2003)}]{lubeck2003B}
\bibinfo{author}{\bibfnamefont{S.}~\bibnamefont{Lubeck}} \bibnamefont{and}
  \bibinfo{author}{\bibfnamefont{P.~C.} \bibnamefont{Heger}},
  \bibinfo{journal}{Phys. Rev. E.} \textbf{\bibinfo{volume}{68}},
  \bibinfo{pages}{056102} (\bibinfo{year}{2003}).

\bibitem[{\citenamefont{Albert and Barab\'asi}(2002)}]{albert2002}
\bibinfo{author}{\bibfnamefont{R.}~\bibnamefont{Albert}} \bibnamefont{and}
  \bibinfo{author}{\bibfnamefont{A.-L.} \bibnamefont{Barab\'asi}},
  \bibinfo{journal}{Rev. Mod. Phys.} \textbf{\bibinfo{volume}{74}},
  \bibinfo{pages}{47} (\bibinfo{year}{2002}).

\bibitem[{\citenamefont{Newman}(2003)}]{newman2003}
\bibinfo{author}{\bibfnamefont{M.~E.~J.} \bibnamefont{Newman}},
  \bibinfo{journal}{SIAM Rev.} \textbf{\bibinfo{volume}{45}},
  \bibinfo{pages}{167} (\bibinfo{year}{2003}).

\bibitem[{\citenamefont{Boccaletti et~al.}(2006)\citenamefont{Boccaletti,
  Latora, Moreno, Chavez, and Hwang}}]{latora2006}
\bibinfo{author}{\bibfnamefont{S.}~\bibnamefont{Boccaletti}},
  \bibinfo{author}{\bibfnamefont{V.}~\bibnamefont{Latora}},
  \bibinfo{author}{\bibfnamefont{Y.}~\bibnamefont{Moreno}},
  \bibinfo{author}{\bibfnamefont{M.}~\bibnamefont{Chavez}}, \bibnamefont{and}
  \bibinfo{author}{\bibfnamefont{D.~U.} \bibnamefont{Hwang}},
  \bibinfo{journal}{Phys. Rep.} \textbf{\bibinfo{volume}{424}},
  \bibinfo{pages}{175} (\bibinfo{year}{2006}).

\bibitem[{\citenamefont{Dorogovtsev et~al.}(2008)\citenamefont{Dorogovtsev,
  Goltsev, and Mendes}}]{mendes2008}
\bibinfo{author}{\bibfnamefont{S.~N.} \bibnamefont{Dorogovtsev}},
  \bibinfo{author}{\bibfnamefont{A.~V.} \bibnamefont{Goltsev}},
  \bibnamefont{and} \bibinfo{author}{\bibfnamefont{J.~F.~F.}
  \bibnamefont{Mendes}}, \bibinfo{journal}{Rev. Mod. Phys.}
  \textbf{\bibinfo{volume}{80}}, \bibinfo{pages}{1275} (\bibinfo{year}{2008}).

\end{thebibliography}

\end{document}